\documentclass[12pt]{article}

\usepackage{amssymb}
\usepackage{amsmath}
\usepackage{epsfig}
\usepackage{graphicx}

\def\be{\begin{equation}}
\def\ee{\end{equation}}
\def\ba{\begin{eqnarray}}
\def\ea{\end{eqnarray}}

\begin{document}
\hskip 1 cm
\vskip 0.5cm

\vspace{25pt}
\begin{center}
    { \LARGE{\bf Towards a gauge invariant volume-weighted\\ \vskip 3mm 
    probability measure for eternal inflation}}
    \vspace{33pt}

  {\large  {\bf   Andrei Linde}}

    \vspace{15pt}

 {Department of Physics,
    Stanford University, Stanford, CA 94305}

  \end{center}

   \vspace{20pt}

\begin{abstract}
An improved volume-weighted probability measure for eternal inflation is proposed. For the models studied in this paper it leads to simple and intuitively expected gauge-invariant results.
\end{abstract}

\vspace{10pt}


\section{Introduction}

Soon after the invention of inflationary cosmology it was realized that inflation may divide our universe into many exponentially large domains corresponding to
different metastable vacuum states, forming a huge inflationary
multiverse \cite{linde1982,nuff}. The total number of
such vacuum states in string theory can be enormously large
\cite{Lerche:1986cx,Bousso:2000xa,Kachru:2003aw,Douglas}. A combination of these two facts with the eternal inflation scenario \cite{linde1982,Vilenkin:1983xq,Eternal} is called the theory of the inflationary multiverse \cite{book,LLM,anthropic}, or the  
string landscape scenario \cite{Susskind:2003kw}. 

In order to make any kind of predictions in the context of this new scientific paradigm, one should find all possible vacua of the theory, study all possible versions of the cosmological evolution which can bring us to each of these vacua, and assign certain probabilities to all possible outcomes of this evolution compatible with the fact of our existence.

This last part of the problem is especially complicated and ambiguous; see \cite{Linde:2006nw,Winitzki:2006rn,Aguirre:2006ak} for a recent discussion. For example, one may try to calculate the probability to be born in a part of the universe with given properties {\it at a given point}, ignoring creation of new volume (new points) during inflation. One can consider a small box of a given size $\Delta x$ in comoving coordinates,  which do not change during the expansion of the universe, take a representative set of points in this box, and follow the evolution of physical conditions (fields, temperature, etc.) at each of these points without rewarding them for the different rate of the growth of volume of the universe near each of the points  \cite{Starobinsky:1986fx,Goncharov:1987ir,Garriga:1997ef,Bousso:2006ev,Linde:2006nw}. 

Unfortunately, the results of the investigation using this set of the probability measures depend on initial conditions.  More importantly, most of the physical entities which could be associated with ``points'' did not even exist before the beginning of inflation: protons did not exist, photons did not exist, galaxies did not exist. They appeared only after inflation, and their total number  is proportional to the growth of volume during inflation. 

This gave rise to a set of different volume-weighted and/or pocket-weighted probability measures \cite{Eternal,Goncharov:1987ir,LLM,Bellido,Mediocr,Vilenkin:1995yd,Linde:1995uf,Vilenkin:2006qf,Garriga:2005av,Easther:2005wi,Linde:2006nw,Vilenkin:2006qg,Aguirre:2006na,volume}. The main problem with this approach is the embarrassment of riches: The total volume of the universe occupied by any particular vacuum state, integrated over the full history of the eternally inflating universe, is infinitely large. Thus we need to compare infinities, which is a very ambiguous task, with the answer depending on the choice of the cut-off procedure.

Of course, this may simply mean that if life is at all possible in any of these vacua, it is  100\% guaranteed to appear in any of such vacua. Therefore, instead of  calculating the probability to be born in any of these vacua, one may try to find out in which family of the vacua do we live, according to our observations. Then we can take these data as an initial input for all subsequent calculations, and study  conditional probabilities for the quantities which we have not measured yet \cite{Linde:2006nw}. This is a standard approach used by experimentalists who re-evaluate the probability of various outcomes of their future experiments on the basis of the latest experimental data.

This  approach  is quite informative despite its limited nature. However, it is very tempting to go beyond this method  and  try to find a preferable way to introduce the cut-off and evaluate the relative probabilities to be born in different vacua. The simplest probability measure proposed in  \cite{Eternal,Goncharov:1987ir,LLM,Bellido} introduces the cut-off at the upper bound of integration with respect to time. Unlike many other measures, this measure does not require different rules for different situations, it does not suffer from the so-called Boltzmann brain problem, and it is well suited for solving the cosmological constant problem  \cite{Linde:2006nw}. Since this cut-off procedure depends on time in an exponentially expanding universe, it gives the results that can be exponentially sensitive to the choice of time parametrization \cite{LLM,Bellido}.  

This observation stimulated the search for other probability measures which are not sensitive to the choice of time parametrization \cite{Vilenkin:1995yd,Linde:1995uf,Vilenkin:2006qf,Garriga:2005av,Easther:2005wi,Vilenkin:2006qg}.  However, the results obtained by using all of these measures remain sensitive to the choice of the cut-off. As the authors of this set of measures emphasized, their proposals allow a certain degree of flexibility which can be used to further specify their predictions; see e.g.  \cite{Garriga:2005av,Vilenkin:2006qg}.  Therefore the search for the best probability measure still continues. 

In this paper we will take a fresh look at the simplest volume-weighted probability measure \cite{Eternal,Goncharov:1987ir,LLM,Bellido}. We will identify the physical reason for the exponential sensitivity of the results to the choice of time parametrization \cite{LLM,Bellido}. After that, we will  propose an improved version of the volume-weighted measure, which does not suffer from this problem, at least in the simple models that we have been able to analyze so far.

\section{Volume-weighted distribution and bubble formation}\label{VW}

The methods that we are going to use in our investigation were described in our previous paper \cite{Linde:2006nw}; here we will only briefly remind the reader of the necessary ingredients, and mention certain limits of applicability of these methods. 

In order to investigate a possible dependence of our results on the choice of the time slicing, we will will study slices of equal time with time measured in units $H^{\beta-1}M_{p}^{1-\beta}$. The   choice $\beta = 1$ corresponds to standard comoving time measured by number of oscillations. The choice $\beta = 0$ would mean that the time is measured in units of $H^{{-1}}$, or, equivalently, in the number of e-foldings. In such a case all parts of the universe expand at the same rate; this corresponds to the pseudo-comoving probability distribution of Ref. \cite{Linde:2006nw}. We are unaware of any physical interpretation of other values of $\beta$; in this paper we will consider $\beta$ in the interval $0\leq \beta \leq 1$ and keep track of the dependence of our results on $\beta$ in order to check whether the final results  depend on time parametrization. 

\begin{figure}[hbt]
\centering
\includegraphics[scale=0.25]{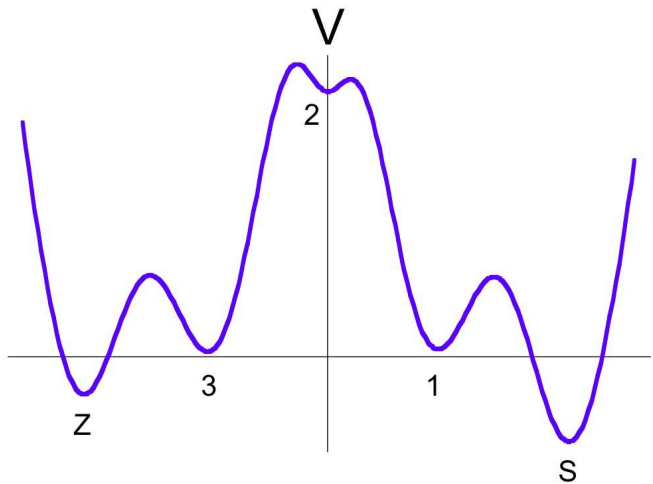} \caption{A potential with three dS minima and two sinks.}\label{2sinksnobrains}
\end{figure}

Consider first the theory with a potential shown in Fig. \ref{2sinksnobrains}.

All $dS_{i}$ states are unstable and decay to other states $dS_{j}$ with the decay rates $\Gamma_{ij}$. Strictly speaking, one should multiply $\Gamma_{ij}$ by an additional factor ${4\pi\over 3}H_{i}^{\beta-4}$  \cite{Vilenkin:2006qf}. However, in this paper we will assume that all factors $\Gamma_{ij}$ are exponentially small and ignore all sub-exponential factors. As we will see in the next section, the main problems that we encounter appear already in this approximation.

 Let us consider, for definiteness, the part of the universe of size $H_{2}^{{-3}}$ in de Sitter state $dS_{2}$ corresponding to the upper local minimum of the potential in Fig. \ref{2sinksnobrains}. (The results of our approach do not depend on this particular choice.) All points inside the box initially are causally connected. Each point of $dS_{2}$ will transit either to $dS_{1}$ or to $dS_{3}$, from where it may either return back or go to a collapsing universe corresponding to $AdS_{s}$ or $AdS_{z}$.  In this paper we will assume that collapsing parts disappear into a singularity and can be ignored for all practical considerations. Therefore we will call  $AdS_{s}$ and $AdS_{z}$ terminal vacua, or sinks. 
  
 Following \cite{Linde:2006nw}, one may begin with the investigation of this process in comoving coordinates, i.e. ignoring the expansion of the universe. To get a visual understanding of the process of bubble formation in comoving coordinates, one may paint red all of the parts of the original box corresponding the state $dS_{2}$, paint blue the parts in the state $dS_{1}$, paint green  the parts in the state $dS_{3}$, and paint black all points in any of the two sinks.  In comoving coordinates, the size and the shape of the original box does not change during the cosmological evolution; all points remain at the same positions (same coordinates), but the values of the fields at each of these points (their colors) may change. 
 
In the absence of the sinks, the whole universe would become divided into red, blue and green parts, with the relative comoving volume of each of the parts proportional to $e^{{\bf S}_{i}}$, where ${\bf S}_{i}$ is the entropy of each of the vacua. On the other hand, in the presence of the sinks, each particular points will eventually end up in the black. However, this does not mean that the whole universe is going to die. Indeed, under our assumption that $\Gamma_{ji}$ are exponentially small, the total volume remaining in each of the dS vacua will continue growing exponentially. This is a crucial feature of the eternal inflation scenario, but it is hard to see it in comoving coordinates, ignoring the growth of volume.

Here we have two different possibilities. If we associate ourselves with some particular point which existed at the moment of creation of the universe, then we can study the most probable evolution of this point during the time until it dies in the sink. One could hope that by doing so we will find the most probable trajectory which could bring us to the present state. This would correspond to the evaluation of comoving coordinate probabilities  \cite{Starobinsky:1986fx,Goncharov:1987ir,Garriga:1997ef,Bousso:2006ev}. 

However, this proposal is rather problematic. Let us assume, for example, that when our universe was born, it was a compact universe of a Planckian size. According to \cite{Linde:1983mx,Linde:2004nz}, quantum creation of such universes seems most probable. This result was recently used in one of the papers advocating the usefulness of comoving probabilities \cite{Bousso:2007er}.
But Planckian-size universes in the beginning could contain just a few physically distinguishable points: We do not know any way to distinguish between  points separated by less than the Planckian distance.

Thus we cannot associate ourselves with anything existing at the beginning of inflation: The total number of physically distinguishable points in the universe at the moment of its creation in the model discussed above is much smaller that the number of people living now on the Earth. A similar conclusion is true for most of the inflationary models independently of their initial state: If we consider any two observers living now on the Earth, the distance between them at the beginning of inflation was smaller than Planckian. Moreover, all elementary particles in our bodies  did not exist before inflation, and even the galaxy where we live did not exist before inflation. Therefore it is hard to imagine that the behavior of the physical conditions {\it at a given initial point}, ignoring production of many new points during the cosmological evolution, could have any anthropic significance. In a certain sense, the eternal  existence of an infinite number of points where life may or may not emerge is just a bad mathematical idealization, which is especially misleading when applied to inflationary cosmology.

One of the main roles of inflation is to create exponentially large number of elementary particles in each part of an inflationary  universe which originally occupied an extremely small causally connected region of the size $O(H^{{-1}})$. But in order to study the new parts of the universe eternally created by inflation one should pay attention to the exponential growth of the volume of the universe. One of the ways to do it is to study the volume-weighted probability distribution introduced in \cite{Eternal,LLM,Bellido}. In application to the scenario with transitions between the metastable dS vacua, the set of equations for the probability density $P_{i}$ for  the potential with three different dS minima and two AdS minima  was given in  \cite{Garriga:1997ef,Linde:2006nw}: 
\begin{equation}
\label{v1vw000} \dot P_{1} =  -P_{1}\, (\Gamma_{1s} +\Gamma_{12}) +  P_{2}\,\Gamma_{21} +3H_{1}^{\beta}P_{1}\ ,
\end{equation}
\begin{equation}
\label{v2vw000} \dot P_{2} =  -P_{2}\, (\Gamma_{2s} +\Gamma_{21}+\Gamma_{23}) +  P_{1}\,\Gamma_{12}+P_{3}\,\Gamma_{32} +3H_{2}^{\beta}P_{2}\ ,
\end{equation}
\begin{equation}
\label{v3vw000} \dot P_{3} =  -P_{3}\, (\Gamma_{3z} +\Gamma_{32}) +  P_{2}\,\Gamma_{23} +3H_{3}^{\beta}P_{3}\ .
\end{equation}
The first few terms in these equations describe the transitions between different vacua. The last terms describe the growth of volume of de Sitter spaces $dS_i$.
These equations intuitively are very simple, but one should keep in mind that their derivation involved several approximations  \cite{Garriga:1997ef}.

We will be interested in the case where $H_{2}^{\beta}$ is much greater than all other parameters in these equations. In this case $P_{2}$ obeys a simple equation
\begin{equation}
\label{v2vw000oo} \dot P_{2} = 3H_{2}^{\beta}P_{2}\ ,
\end{equation}
i.e. in the first approximation $P_{2}$ does not depend on $P_{1}$, $P_{3}$:
\be
P_{2} = P_{2}(0)\, e^{3H_{2}^{\beta}\, t} \ .
\ee
Because of the fast growth of $P_{2}$, the terms $P_{2} \Gamma_{2i}$ eventually become the leading terms in the equations for $P_{i}$, for all $i\not = 2$:
\begin{equation}
\label{v1vw000i} \dot P_{i}  =  P_{2}\,\Gamma_{2i} \ .
\end{equation}
The solution of this equation is
\begin{equation}
\label{v1vw000i2}  P_{i}(t) =  {P_{2}(t) \,{\Gamma_{2i}}\over 3H_2^{\beta}}~  \left(1-e^{-3H_{2}^{\beta}\, t}\right) \ .
\end{equation}
Within a very short time $O(H_{2}^{-\beta})$ the solution approaches the asymptotic regime
\begin{equation}
\label{v1vw000i2}  P_{i}(t) =  {P_{2}(t) \,{\Gamma_{2i}}\over 3H_2^{\beta}} =  {P_{2}(0) \,{\Gamma_{2i}}\over 3H_2^{\beta}}~ e^{3H_{2}^{\beta}\, t}\ .
\end{equation}
The ratio of $P_i$ is given by
\be\label{tunndown}
p_{ij} = {P_{i}\over P_{j}} ={\Gamma_{2i}\over \Gamma_{2j}} \ , 
\ee
where $i,j \not = 2$. Note that this result {\it does not} depend on the time parametrization described by the factor $\beta$ in the approximation specified above.

Applying these results to anthropic considerations requires several additional steps.  The functions $P_{i}$ describe the total volume of all parts of the universe in a given vacuum state,  produced from the initial domain. However, when the bubbles of a new phase expand, their interior eventually becomes an empty dS space. If we are  usual observers born after reheating of the inflationary universe, then one may argue that the probability to be born in the bubble $dS_{i}$ during any given time interval is proportional not to the total volume of de Sitter bubbles, which is described by $P_i$, but to its part produced within some finite time after the bubble formation, e.g. in the interval from 1 to 100 billion years. In other words, the probability of life is proportional not to the total volume, but to the new volume, which switches our attention form the total probability $P_i$ to the {\it incoming} probability charge $Q_i$ and the incoming probability current $\dot Q_i$ \cite{Bellido,Garriga:1997ef,Bousso:2006ev}. When we say `incoming,' we  emphasize that we are interested only in the growth of the volume of the new parts of the bubbles, but not in the speed of their decay. This growth occurs due to the new bubble production, and due to the expansion of the walls of the existing bubbles. In the case where $H_{2}^{\beta}$ is much greater than all other $H_i^\beta$, the leading part of this process occurs due to the new bubble production, and equations for $Q_i$ have the following simple form, coinciding with the form of the leading terms in the equations for $P_i$:
\begin{equation}
\label{v3vw000q} \dot Q_{1} =  P_{2}\,\Gamma_{21}  \ .
\end{equation}
\begin{equation}
\label{v3vw000q} \dot Q_{2} =     P_{1}\,\Gamma_{12} +P_{3}\,\Gamma_{32}  \ .
\end{equation}
\begin{equation}
\label{v3vw000q} \dot Q_{3} =    P_{2}\,\Gamma_{23}  \ .
\end{equation}

In this case, the ratio of the probability currents from the upper minimum to the lower minima will be given by
\be\label{tunndown}
q_{ij} = {Q_{i}\over Q_{j}} = {\dot Q_{i}\over \dot Q_{j}} = {P_{i}\over P_{j}} ={\Gamma_{2i}\over \Gamma_{2j}} \ , 
\ee
for $i,j \not = 2$. As before, this result {\it does not} depend on the time parametrization described by the factor $\beta$ in the approximation specified above.

Thus,  the predictions obtained using the volume-weighted probability distribution do not depend on the choice of time slicing for a wide variety of time parametrizations. This dependence appears only for exponentially small $\beta$, i.e. in the limit when one completely ignores the expansion of the universe. However, in this limit our equations should be written  more precisely.

Our results do not depend on initial conditions either. Indeed, if we assume that initially the field was positioned not in the highest minimum,  the universe will expand at a slower rate until the jump to the highest minimum occurs. Starting from this time, the volume of the part of the universe in the highest minimum (and the total volume of its descendants produced by its decay)  will grow much faster than all other parts. Eventually this will result in the stationary regime discussed above, with $q_{ij} = {Q_{i}\over Q_{j}} = {\dot Q_{i}\over \dot Q_{j}} = {P_{i}\over P_{j}} ={\Gamma_{2i}\over \Gamma_{2j}}$.

Let us describe  our approximation again. First of all, we assumed that $H_{2}^{\beta}$ is much greater than all decay rates $\Gamma_{ij}$. This is a legitimate approximation as long as we ignore all sub-exponential factors.   More importantly, we assumed that $H_{2}^{\beta}$ is much greater than all other  $H_{i}^{\beta}$. If $H_{2} \gg H_{i}$ for $i \not = 2$, the condition $H_{2}^{\beta} \gg H_{i}^{\beta}$ is violated only for $\beta< \log^{-1} {H_2\over H_i} \ll 1$. We assumed also that as soon as  bubbles of the vacuum $dS_{i}$ form, their interiors start growing exponentially, as $e^{3H_{i}^{\beta} t}$, and starting from the moment of their formation,  decays to other vacua with the decay rate $\Gamma_{ij}$ become possible.

This approximation seems reasonable, but one should remember that it is just an approximation. For example, if one restores the sub-exponential term ${4\pi\over 3}H_{i}^{\beta-4}$ in the definition of $\Gamma_{ij}$, one will encounter the sub-exponential dependence of the results on $\beta$. It may happen that this dependence will disappear when one solves all other parts of the problem without making any approximations. We will not discuss this issue here since we have a much more complicated problem to solve:  Once we add to this picture a stage of  slow-roll inflation, the dependence on $\beta$ becomes exponential \cite{Bellido}. In the next section we will discuss this problem, identify its source, and propose its possible resolution.

\section{Volume-weighted measure for slow-roll inflation}\label{slow}

\begin{figure}[hbt]
\centering
\includegraphics[scale=0.30]{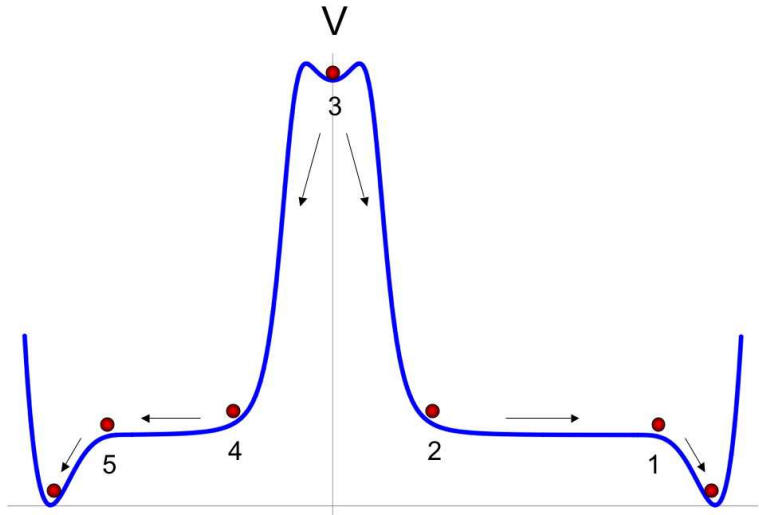} \caption{Tunneling from the false vacuum to two different slow roll inflation regimes.}\label{slownum}
\end{figure}

Let us now consider inflation in a theory with the potential shown in Fig. \ref{slownum}, which describes both the tunneling and the slow-roll. Inflation begins while the field $\phi$ is at the top of the potential, $\phi = \phi_{3}$. Then the field tunnels either to $\phi_{4}$ or to $\phi_{2}$, and experiences a slow-roll regime until it reaches either $\phi_{5}$ or $\phi_{1}$, which corresponds to the end of inflation boundary. Once this happens, the field  instantly falls into the minimum nearby, and the corresponding part of the universe becomes hot.

The first part of this process can be described by the equations similar to the ones studied in the previous section, with an obvious result: At $ H_{3}^{\beta} t \gg1$ the probability distributions rapidly approach an asymptotic regime
\be
P_{3}(t) \sim  P_{3}(0)\ e^{3H_{3}^{\beta}t}, \qquad P_{4} =Q_{4} = P_{3}(0)\   \Gamma_{34}e^{3H_{3}^{\beta}t}, \qquad    P_{2}= Q_{2} = P_{3}(0)\ \Gamma_{32}e^{3H_{3}^{\beta}t} \ .
\ee
Just as before, the ratio $Q_{4}/Q_{2} = \dot Q_{4}/\dot Q_{2} = P_{4}/P_{2} = \Gamma_{34}/\Gamma_{32}$ does not depend on time $t$ and on the time parametrization $\beta$, which is encouraging. 

However, let us now discuss the subsequent slow-roll stage. This stage cannot be described by the simple equations used in the previous section. Equations for the tunneling allowed almost instant response for the exponential growth of the probability distribution $P_{2}$: Once the total volume of the universe at the top of the potential starts growing exponentially as $e^{3H_{2}t}$, this growth almost instantly results in the proportional growth of all other $P_{i}$. The situation during the slow-roll regime is qualitatively different: The field cannot tunnel anywhere, and the universe cannot reheat until the slow-roll regime is over. One can describe this effect using the volume-weighed Fokker-Planck equations \cite{LLM}, or using a simplified slow-roll description ignoring quantum diffusion of the scalar field $\phi$ on its way down. Here we will use this simplified approach; it will be quite adequate for our purposes.

After tunneling to the point $\phi_{4}$, the field $\phi$  slowly rolls to the left  during some time $t_{45}$ until it reaches the point $\phi_{5}$, where inflation ends. Suppose for simplicity that during this time the universe inflates with the Hubble constant $H_{45} \approx const$.  Similarly, the field $\phi$  rolls to the right from $\phi_{2}$ during some time $t_{21}$ until it reaches the point $\phi_{1}$ where inflation ends. During this time the universe inflates with the Hubble constant $H_{21} \approx const$. We will assume, for definiteness, that $t_{21} > t_{45}$.

Because of the rapid growth of the upper dS space, the total volume of all parts of the universe eventually approaches a stationary regime, when the total volume of all parts of the universe will grow approximately as $e^{3H_{2}t}$. This is a very counter-intuitive result, which we already discussed in the previous section, but now we need to understand it in a much more detailed way.

Let us consider this process step by step. Suppose the tunneling begins at the moment $t = 0$. The number of bubbles with the field $\phi_{2}$ at their center will be the same as the number of the bubbles with the field $\phi_{4}$ at their center. (Note that in the coordinate system that we are using the bubbles have finite size, and the initial value of the field at their center is determined by the CDL tunneling trajectory.) But during the time interval $t_{21}$ after the tunneling there will be no bubbles with the field $\phi_{1}$ at their center. When they  appear, their total volume will be equal to the total volume of the first bubbles with the field $\phi_{2}$, multiplied by $e^{3N_{21}}$, where $N_{21} = H_{21} t_{21}$ is the number of e-folds of the slow-roll inflation on the right branch of the potential. Note that this factor does not depend on $\beta$. The factor $e^{3N_{21}}$ describes the total growth of volume during the slow-roll inflation; it is gauge invariant.  

Starting from  $t = t_{21}$, the total volume occupied by the field $\phi_{1}$  grows as $e^{3H_{3}(t - t_{21})}$:
\ba\label{right}
P(\phi_{1}, t) &=&0\hskip 5.4cm {\rm for}~ t< t_{21} \ ,\nonumber \\
P(\phi_{1}, t) &=& P_{3}(0)\   \Gamma_{32}\ e^{3N_{21}}\ e^{3H_{3}^{\beta}(t - t_{21})}\qquad {\rm for}~ t> t_{21} \ .
\ea
Similarly, for the left branch we have 
 \ba\label{left}
P(\phi_{5}, t) &=&0\hskip 5.4cm {\rm for}~ t< t_{45} \ ,\nonumber \\
P(\phi_{5}, t) &=& P_{3}(0)\   \Gamma_{34}\ e^{3N_{45}}\ e^{3H_{3}^{\beta}(t - t_{45})}\qquad {\rm for}~ t> t_{45} \ .
\ea
If we literally follow the rules of \cite{LLM}, we should divide $P(\phi_{1}, t) $ by $P(\phi_{5}, t)$ at large $t$, to obtain the following result:
\be
{P(\phi_{1}, t)\over P(\phi_{5}, t) }  =   {\Gamma_{32}\ e^{3N_{21}}\over  \Gamma_{34}\ e^{3N_{45}}}\ e^{3H_{3}^{\beta}\Delta t} \ .
\ee
Here
\be
\Delta t = t_{45} - t_{21} \ .
\ee
This result is exponentially sensitive to the time parametrization factor $\beta$ \cite{LLM}. This was the  main fact that forced many authors to look for a better probability measure for eternal inflation.

However, in the simple example considered above one can easily identify the source of the problem. We were able to find a time-independent result for the ratio of the two probabilities because both of the expressions 
$P(\phi_{1}, t)$ and  $P(\phi_{5}, t)$ {\it asymptotically} approach the same stationary regime; they grow at the same (exponential) rate $e^{3H_{2}^{\beta}t}$. But they approach the stationary regime at two {\it different } moments of time because of the time delay $\Delta t = t_{45} - t_{21}$ related to the slow roll regime.

\begin{figure}[hbt]

\

\centering
\includegraphics[scale=0.28]{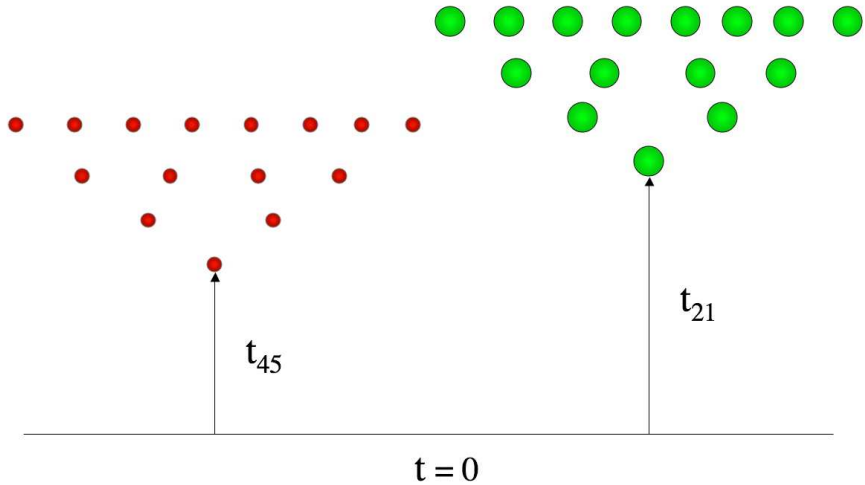} 

\caption{Large green bubbles correspond to the bubbles containing the field $\phi_1$. Small red bubbles contain field $\phi_5$.  The chain reaction of the bubble production starts at the time $t_{12}$ for the `tree' of the bubbles of the field $\phi_1$, and at the time $t_{45}$ for the `tree' of the bubbles of the field $\phi_5$. After this process starts, the number of the  bubbles belonging to each `tree' grows at the same rate, as $e^{3H_2\tau}$, where $\tau$ is the time starting from the moment when the first bubble of each type appears. Instead of comparing the number of the bubbles at the same moment of time $t$  \cite{LLM}, which does not make much sense for $t < max\{t_{21}, t_{45}\}$, we propose to reset the time for each type of the bubbles to the moment $\tau = 0$, when their number starts growing at the same rate, and compare them at the same time $\tau$. In other words, one should compare apples to apples, instead of comparing apples to the trunks of the trees. The results of this comparison in our simple model do not depend on initial conditions, on the time $\tau$,  or  on the choice of time parametrization.}\label{colortrees}
\end{figure}

In the context of this scenario, it does not make much sense to compare two time-dependent quantities before they have reached the asymptotic stationary regimes. This problem was hidden in the original proposal of  \cite{LLM,Bellido} because we studied there the limit  $t\to \infty$, where this issue could not be clearly identified. Therefore we believe that our original proposal to compare different processes at the same moment of time, {\it starting from the beginning of inflation}, is incomplete and requires a certain modification. Indeed, the probability distribution proposed in  \cite{LLM,Bellido}  was very far from stationary at the beginning of the process, when $t < max\{t_{21}, t_{45}\}$, see Fig. \ref{colortrees}. The simplest modification is to compare the  processes in different vacua (or at different values of the scalar fields, or temperature, or density) {\it starting at the time when the stationary regime becomes established in these states.} The corresponding moments of time will be different for the vacua of each particular type. We will call the improved  probability measure {\it stationary}, to distinguish it from the {\it asymptotically stationary} measure proposed in \cite{LLM,Bellido}.

This means that we should {\it reset our clock}, ignoring the time delay related to the different onset of stationarity for the parts of the universe with different properties. In the example considered above, this means that we should compare the volume at the point $\phi_1$  at the time $\tau = t -  t_{21}$ with   the volume at the point $\phi_5$  at the time $\tau = t -  t_{45}$.  This leads to the improved expression for the ratio of the two probabilities:
\be\label{result}
{P(\phi_{1}, \tau)\over P(\phi_{5}, \tau) }  =  {\Gamma_{32}\ e^{3N_{21}}\over  \Gamma_{34}\ e^{3N_{45}}} \ .
\ee
This is the main result of our paper. This result is gauge-invariant, it does not depend on $\beta$, unless one takes $\beta \ll 1$, which violates the validity of our approximation. This result has a very simple intuitive meaning: It shows that the fraction of the volume of the universe in a given state is proportional to the probability $\Gamma_{3i}$ to go in a given direction,  and to the total growth of volume along the corresponding inflationary trajectory.

Note, that in the derivation of this result we used an assumption that the evolution of the probability distribution is dominated by the expansion of the domain with the highest energy density. This is indeed the case for all but exponentially small $\beta$. However, this condition is not satisfied in the limit $\beta = 0$, which would correspond to the comoving or pseudo-comoving probability distribution \cite{Linde:2006nw}. In other words, the results of our calculations in the simple model discussed above do not depend on time parametrization for all $\beta$ except for the limiting case of $\beta = 0$. 

This resembles the discussion of the gauge-noninvariance of the calculation of the critical temperature in the unitary gauge as compared to (almost) all other gauges \cite{Kirzhnits:1976ts}. The difference appears there because of the use of non-invariant quantities, such as the scalar potential, and also because of the bad infrared behavior of the higher-loop corrections near the critical point in the unitary gauge. Eventually we have learned how to calculate the critical temperature in a gauge-invariant way, but the corresponding procedure is quite involved, so the original simple but slightly non-invariant methods still remain valuable and widely used. 

A possible reason for the particular behavior of the probabilities calculated in the limit $\beta \to 0$ is that for all sufficiently large $\beta>0$, the time that is required for reaching the stationarity regime depends on the parameters of the theory logarithmically, the qualitative picture of the flow from the upper minimum remains correct, and the calculation of the delay $\Delta t$ does not change. Meanwhile for $\beta \leq 0$ the probability current would flow from the lowest dS vacuum, and the delay time $\Delta t$ would be exponentially large. Using the analogy with the theory of phase transitions, one would argue that one should calculate the probabilities by our method only if $\beta$ is not too small; the calculations in the limit $\beta = 0$ using our method become unreliable. Admittedly, this is just an argument; this issue requires a more detailed analysis.  

Let us formulate our results in a different form. Suppose we are interested in the relative volumes of the parts of the universe with two different temperatures $T_{1}$ and $T_{2}$ in the two different vacua. The generalization of our previous result is obvious:
\be
{P(T_{1}, \tau)\over P(T_{2}, \tau) }  =  {\Gamma_{32}\ {\cal{V}}(T_{1})\over  \Gamma_{34}\ {\cal{V}}(T_{2})} \ .
\ee
Here $ {\cal{V}}(T_{2})$ is the total growth of volume of the universe from the moment of tunneling to the point $\phi_{2}$ to the moment when inflation ends and  the temperature in the corresponding part of the universe becomes equal to $T_{1}$.
Similarly, $ {\cal{V}}(T_{2})$ is the total growth of volume of the universe from the moment of tunneling to the point $\phi_{4}$ to the moment when inflation ends and  the temperature in the corresponding part of the universe becomes equal to $T_{2}$.

If we would like to compare the volumes for the same vacuum state, at different temperatures, we would get a trivial result
\be
{P(T_{1}, \tau)\over P(T_{2}, \tau) }  =  {{\cal{V}}(T_{1})\over  {\cal{V}}(T_{2})} \ .
\ee

It is instructive to compare this result to the result which we would obtain using the old measure of \cite{LLM,Bellido} prior to its improvement: 
\be
{P(T_{1}, \tau)\over P(T_{2}, \tau) }  =  {{\cal{V}}(T_{1})\over  {\cal{V}}(T_{2})}\ e^{H_{3}^{\beta}\Delta t} \ ,
\ee
where $\Delta t = t(T_{2}) - t(T_{1})$, and $t(T)$ is the time when the universe cools down to the temperature $T$. Clearly, our old measure  would imply exponential dominance of the high temperature regions \cite{Tegmark:2004qd}, which was the essence of the so-called youngness paradox \cite{Guth:2007ng}. However, once we reset our clock so that the comparison begins starting from the time when the stationary regime is established, the dangerous factor $e^{H_{3}^{\beta}\Delta t}$ leading to the  youngness paradox disappears.

One should note that this was indeed a paradox rather than a problem, since the volume distribution obtained in \cite{LLM} was not supposed to be used in anthropic considerations. Meanwhile the volume-weighted anthropic probability measure, which was first proposed in \cite{Bellido}, uses the methods developed in \cite{LLM}, but calculates the probability currents through the hypersurface of a given temperature/density/scalar field. With the probability measure proposed in  \cite{Bellido}, the youngness paradox does not appear \cite{Linde:2006nw}.  (This fact remained unnoticed in many subsequent discussions of the volume-weighted probability measure, see e.g.  \cite{Tegmark:2004qd,Guth:2007ng}.) However, it is quite encouraging that with our improved prescription for  volume-weighted probabilities,  the youngness paradox disappears altogether, even if one decides to compare completely different regions of the universe, without restricting this comparison to the hypersurface of the end of inflation, or of a given temperature or density.

\newpage

\section{Discussion}\label{disc}

In this paper we proposed an improved volume-weighted probability distribution for the multiverse. We performed our investigation ignoring sub-leading factors. For the simple model described in Sections \ref{VW} and \ref{slow}, our approach produced gauge-invariant results, thus avoiding the problem plaguing the probability measure proposed in \cite{LLM,Bellido}. The results obtained by using  our improved volume-weighted measure, which we called {\it stationary}, have a simple intuitive interpretation, which shows that  probability is proportional to the growth of volume during the slow-roll inflation.

The main reason for the gauge invariance of our result can be understood as follows. In any time parametrization, the probability distribution  becomes time-independent once the asymptotic stationary regime is reached. In our model the post-inflationary stationarity is established when the field inside the bubbles produced by the tunneling from the state $dS_{3}$ reaches the end of inflation (points 1 and 5). Since the definition of the end of inflation is gauge-invariant, the probability distribution at the corresponding hypersurface is also gauge-invariant, and it remains gauge invariant at all subsequent time slices because of the stationarity.

It is interesting to compare the results of our approach with the results which could be obtained by other methods. At the first glance, our approach may resemble the method developed in \cite{Vilenkin:1995yd}, where the cut-off procedure depended on the fraction of the thermalized comoving volume. However, these methods are quite different. In particular, our results, unlike the results of \cite{Vilenkin:1995yd}, do not depend depend on initial conditions.

  Our main result for the probability distribution (\ref{result}) in the simple model   with the potential shown in Fig. \ref{slownum} coincides   with the result obtained by Vilenkin {\it et al} using a  different probability measure developed in \cite{Vilenkin:2006qf,Garriga:2005av}. However, this coincidence  disappears when one studies more complicated models. For example, one may consider a potential with an additional de  Sitter minimum $dS_{6}$, somewhere on the way from the point 2 to point 1. As one may anticipate from the results obtained in the previous section, this should lead to an additional suppression of the probability to live in the minimum near the point 1 by the exponentially small factor $\Gamma_{61}$.  The meaning of this result is very simple: The exponentially growing probability current flows from the upper vacuum in the direction of the least resistance. When we add an extra barrier, the corresponding current slows down. This conclusion can be easily verified by solving the corresponding system of equations. 
  
In the similar situation studied in \cite{Garriga:2005av} the authors identified two possible results of the introduction of the additional minimum. The first of the two probability measures discussed in  \cite{Garriga:2005av} predicts that the insertion of an additional metastable minimum $dS_{6}$ does not affect the ratio of the probabilities to live in the right minimum as compared to the left minimum. Meanwhile the second of these two measures, which was advocated in  \cite{Garriga:2005av}, suggests that the insertion of the additional minimum makes the relative probability to live in  the left minimum vanishingly small. This is directly opposite to the effect expected in our approach. Note that this difference appears even in the models without the slow-roll regime.

This conclusion, however, requires a more detailed investigation. The calculation of the time delay  for the simple model shown in Fig. \ref{slownum} was quite unambiguous. In more complicated models with many minima it may be much more difficult to identify the time when the stationary regime is established. A small error in determining this time may translate into a very large error in the relation between the probabilities, proportional to $e^{3H^\beta_2 \Delta t}$. We did not find any such terms in the models without the slow-roll regime, but it is quite possible that a more accurate investigation will reveal them. 

The situation may become even more interesting and complicated if one considers Hawking-Moss tunneling, or includes the stages of eternal slow-roll inflation. This will require an investigation of the branching diffusion  equations along the lines of \cite{LLM,Bellido}. Thus the calculation of the probabilities in the landscape can be a rather challenging task. But nobody expected an investigation of the eternally growing inflationary multiverse to be simple. Instead of complaining about the hardship of life, let us try to concentrate on the positive news.

There are many different proposals for the calculation of the volume-weighted/pocket-weighted probabilities during eternal inflation. Now we reached the point where the most popular proposals for the volume-weighted or pocket-weighted probability measure give identical results at least for the simple model studied in our paper. This result is quite simple and intuitively clear: The prior probability to live in a given dS state is proportional to the probability of its formation from the dS vacuum with the highest energy density, multiplied by the growth of volume during the slow-roll inflation.

\

\leftline{\bf Acknowledgments}

I am grateful to A. Aguirre, T. Clifton, J. Garriga, A. Guth, L. Kofman, M.~Sasaki, 
S.~Shenker, N.~Sivanandam, L. Susskind, V. Vanchurin, A.~Vilenkin and  S.~Winitzki for valuable discussions. This work was supported by NSF grant
PHY-0244728.  

\

\end{document}